\documentclass[sigconf,authorversion]{acmart}

\AtBeginDocument{%
  }

\setcopyright{none}
\acmConference[Preprint]{}{2026}{}
\usepackage{booktabs}
\usepackage{multirow}
\usepackage{subcaption}

\newcommand{\added}[1]{#1}
\newcommand{\deleted}[1]{}

\begin{document}

\title{Punchlines Unbound: Comedy Practices in Social Virtual Reality}

\author{Ryo Ohara}
\affiliation{%
  \institution{The University of Tokyo}
    \city{Tokyo}
  \country{Japan}}
\email{rohara@cyber.t.u-tokyo.ac.jp}

\author{Chi-Lan Yang}
\affiliation{%
  \institution{The University of Tokyo}
    \city{Tokyo}
  \country{Japan}}
\email{chilan.yang@cyber.t.u-tokyo.ac.jp}

\author{Yuji Hatada}
\affiliation{%
  \institution{The University of Tokyo}
    \city{Tokyo}
  \country{Japan}}
\email{hatada@cyber.t.u-tokyo.ac.jp}

\author{Takuji Narumi}
\affiliation{%
  \institution{The University of Tokyo}
    \city{Tokyo}
  \country{Japan}}
\email{narumi@cyber.t.u-tokyo.ac.jp}

\author{Hideaki Kuzuoka}
\affiliation{%
  \institution{The University of Tokyo}
    \city{Tokyo}
  \country{Japan}}
\email{kuzuoka@cyber.t.u-tokyo.ac.jp}
\renewcommand{\shortauthors}{Ohara et al.}

\begin{abstract}
Social VR platforms serve as an emergent venue for live performance, enabling co-presence and real-time interaction among distributed performers and audiences within shared virtual environments.
Live performances, such as comedy, rely on subtle social cues between performers and audiences, which are missing in VR.
However, it remains unclear how comedians utilize avatar-mediated cues in social VR.
We conducted semi-structured interviews and observations with 23 virtual comedians on VRChat.
Results revealed that virtual comedians transformed their limited nonverbal expressiveness into performative opportunities through intentional control and exaggeration.
Additionally, a distinctive culture emerged around context-appropriate emoji reactions from audiences, while challenges such as audio latency and moderation against trolling were highlighted.
Our findings advance understanding of how performers creatively adapt to expressive constraints in avatar-mediated settings.
We further demonstrate how challenges in performer-audience interaction and moderation provide design insights for systems enhancing feedback visibility and sustain community norms without restricting creative expression.

\end{abstract}


\begin{teaserfigure}
  \includegraphics[width=\textwidth]{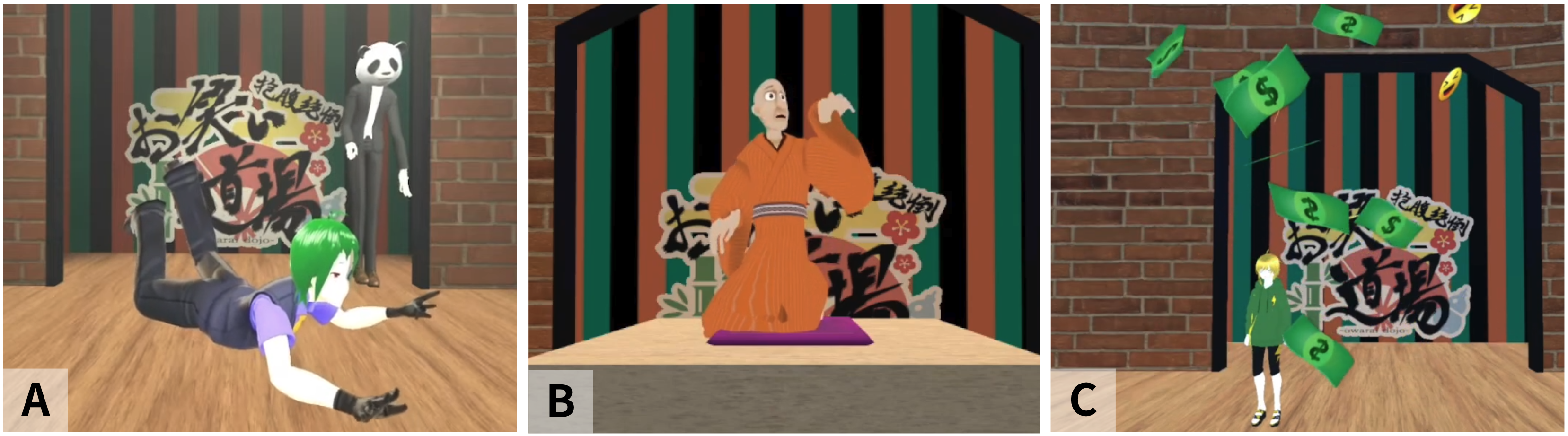}
  \caption{Examples of comedy practices in social virtual reality.  (A) Performers exaggerate avatar movements, such as drifting in space, to compensate for missing subtle nonverbal cues. (B) A performer deliberately controls facial expressions to overcome limited expressivity and convey emotional nuance. (C) Audiences use context-appropriate emoji reactions to provide feedback that complements physical laughter cues while remaining distinct from them.
}
  \label{fig:teaser}
\end{teaserfigure}


\maketitle

\section{INTRODUCTION}
Social virtual reality (social VR) has attracted increasing attention as an emergent venue for virtual live performance. 
It provides a shared environment in which multiple distributed users can interact with one another~\cite{perry2016socialvr}, typically through embodied avatars that represent their presence in the virtual space~\cite{freeman2021socialvravatar}.
This environment offers live performers an alternative virtual stage where they can perform and engage with remote audiences in real-time.
These possibilities have been explored in events such as Sanrio Virtual Festival\footnote{Sanrio Virtual Festival 2025 \url{https://v-fes.sanrio.co.jp/en}} and Intermundium\footnote{Sensorium Galaxy INTERMUNDIUM Full Show Archive \url{https://www.youtube.com/watch?v=YlTBRkurWAE}}, where avatar-mediated live performances demonstrated the creative potential of social VR platforms.
This stands in contrast to video-based live streaming platforms, which often lack co-presence and have been repeatedly criticized for limiting audience engagement~\cite{webb2016distributed,lc2023contradiction}.

However, it is not difficult to imagine that live performance in social VR poses unique challenges.
In general, in physical live performances, performers rely on various social cues, not only from other co-performers but also from the audience, to enable them to improvise and adjust their delivery.
Live performance requires a wide range of social cues to guide improvisation based on audience responses~\cite{webb2016distributed}, including laughter, applause, shifts in collective attention, or moments of silence. 
Prior work has noted that avatars in social VR often fail to convey fine-grained nonverbal cues such as gaze, gesture, and facial expression~\cite{maloney2020socialvrnonverbalcscw,tonenbaum2020socialnonverbalchi}.
This line of research motivates us to explore how performers assess their co-performers and audiences in social VR, where the social cues are all mediated through avatars and virtual environments.
Understanding this will advance our knowledge of how improvisation and audience interaction are sustained under technological constraints, offering implications for the design of live performance in immersive platforms.

Among various forms of live performance, this study focuses on the practice of comedy in social VR, a genre that particularly depends on both creative avatar reappropriation and dynamic performer–audience interaction.
In physical live comedy, performers rely on rich social cues to shape delivery and timing.
For example, gestures and facial expressions play a central role in projecting humor~\cite{scarpetta2009interactionalcomedy,rutter2001rhetoric,tabacaru2014facialhumor}.
In social VR, subtle nonverbal cues are complex to convey~\cite{maloney2020socialvrnonverbalcscw,tonenbaum2020socialnonverbalchi}, while performers also have the advantage of freely designing and controlling their avatars~\cite{freeman2021socialvravatar,nemu2023socialvr}.
However, it remains unclear how comedians balance the constraints of limited nonverbal expressivity with the creative possibilities of avatar design when performing in social VR.
This leads to our first research question, \textit{What \textbf{strategies} do virtual comedians employ to conduct comedy performances in avatar-mediated social VR environments?} \textbf{(RQ1)}
Studying comedy, therefore, provides an opportunity to understand how performers develop strategies to adapt their live performances to the medium's distinctive constraints and possibilities.

Building on the expressive strategies discussed above, live comedy relies not only on prepared material but also on the performers' ability to coordinate with audiences in real-time.
In physical live performance, comedians actively monitor audience reactions and adjust their delivery to sustain comedic flow~\cite{wells2007audiencereaction,harbidge2011audienceship}
Timing is often managed through techniques such as confirmatory utterances before a punchline or strategic pauses that accommodate audience laughter~\cite{rutter1997standup,rutter2001rhetoric,mcilvenny1992laughtraps}. 
These adjustments enable performers to maintain rapport and effectively navigate unpredictable audience responses.
While such interactional responsiveness is a well-established feature of physical comedy settings, it remains unclear how similar forms of coordination are achieved in social VR, where responses must be interpreted through digitally mediated behaviors. 
Thus, we form the second research question: \textit{How does the cueless environment in social VR influence the \textbf{interactions} between performers and audiences?} \textbf{(RQ2)}

Extending beyond individual performance and immediate interaction, sustaining comedy in social VR also requires managing the broader conditions that allow live events to take place.
Unlike physical venues with formal boundaries and staff oversight, social VR platforms enable users to move, speak, and interact without many physical constraints.
While this openness encourages creative engagement, it also makes performances vulnerable to interruptions. Previous studies have reported recurring issues such as trolling and harassment in social VR~\cite{chen2025moderation,chen2025socialvrblocking}, which have prompted the introduction of mechanisms like vote kicking and Trust Rank, a reputation-based system that restricts user permissions based on prior behavior~\cite{chen2024trustsystem}.
However, these mechanisms focus primarily on individual misconduct, and less is known about how participants themselves establish \added{community practices} to manage disruptions and maintain the performative context in social VR.
In comedy settings, where performers and audiences have different goals, it is important to understand the challenges faced by each party and the actions they take to foster sustained participation in shared virtual environments.
Hence, in the third research question, we ask: \textit{What \added{\textbf{community practices}} emerge within the spaces of comedy performance in social VR?} \textbf{(RQ3)} 



Through interviews with 23 virtual comedians in Japan and observations of their performances in VRChat, this study investigated the practice of virtual comedy in social VR.
Our findings revealed that performers reframed the constraints of limited nonverbal expressiveness as opportunities, often exaggerating or intentionally controlling avatar movements to enhance timing and delivery.
Audiences, in turn, developed distinctive practices such as context-adaptive emoji reactions, which opened up a new form of stage feedback for engaging performance in social VR.
\added{Our findings showed that sustainable life performance in social VR requires virtual comedians and audiences to cultivate community practices for managing disruptions, such as trolling or misaligned participation.}

This study extends our understanding of creative practices in social VR by empirically studying how performers and audiences engage in virtual comedy. Based on the insights gained, we discuss ways to design interactions that can better support expressive performances, capture spontaneous feedback from audiences, and implement adaptive moderation tailored to live events in social VR.

\section{RELATED WORK}
\subsection{Social VR}
Social VR is a three-dimensional virtual environment in which multiple remote users experience co-presence through avatars and engage in interactions approximating face-to-face encounters~\cite{perry2016socialvr,mcveigh2018socialvr}.
Representative platforms include VRChat\footnote{A social VR platform. \url{https://hello.vrchat.com}}, Meta Horizon Worlds\footnote{A social VR platform. \url{https://horizon.meta.com/}}, and Cluster\footnote{A social VR platform. \url{https://cluster.mu/en/}}.
A defining aspect of Social VR is the use of embodied avatars~\cite{freeman2021socialvravatar}, which convey users' head and hand movements in real-time via consumer HMDs (e.g., Meta Quest).  
These gestures and postures serve as social cues, supporting dynamics such as turn-taking and attention management.  
These gestures and postures serve as social cues that distinguish Social VR from text- or video-based platforms, affording a more spatial and embodied sense of interaction.  

Building on these embodied features, Social VR has been shown to foster prosocial practices~\cite{maloney2020socialvr,mcveigh2019vrsocialinteraction,hide2025friendship}.  
Prior studies indicate that the co-presence enabled by avatars encourages everyday practices, including shared video co-viewing~\cite{simone2019socialvrvideo}, co-sleeping~\cite{yin2023socialvrsleep,maloney2020socialvrsleep}, and social drinking~\cite{chen2024socialvrdrink}.
Beyond these casual activities, social VR has also become a stage for performative practices such as dance~\cite{piitulainen2022dance} and opera~\cite{striner2021opera}, which together demonstrate that it serves not merely as a communication tool but as an environment for novel forms of collective experience and cultural production.

At the same time, these possibilities are fundamentally shaped by technical and social constraints.  
A persistent challenge is the limited fidelity of nonverbal cues: avatars cannot fully convey subtle expressions such as micro-expressions or gaze direction~\cite{maloney2020socialvrnonverbalcscw,tonenbaum2020socialnonverbalchi}.
\added{While extensive research has been conducted on the design of non-verbal cues for avatars to enhance communication in social VR (e.g., \cite{tonenbaum2020socialnonverbalchi,maloney2020socialvrnonverbalcscw}), the impact of their absence on live performances that rely heavily on non-verbal coordination remains unclear. It is also unclear how performers adapt to this challenge to ensure the quality of their performance and interaction with their co-performers.
Our work focused on comedy, a genre of live performance known for its reliance on timing, improvisation, and audience interaction, to understand the challenges and opportunities of avatar-mediated performance in Social VR.
}


\subsection{Avatar-Mediated Social Interaction}
\added{In virtual environments, avatars were introduced as a means of self-representation that enables interaction~\cite{stephenson1994snow} and have since become a central topic in online entertainment research.}
Prior work shows that users would manipulate their identities through avatars~\cite{birk2016avatar,looy2010playeridentification,nick2006avatarmotivation} and manage relationships with others~\cite{jamie2015avatarrelationship,gale2016doppeleganger,mazalek2009avatarmovements}.
Studies of virtual worlds such as Second Life report practices including marriage, education, and business conducted via avatars, positioning avatars not merely as playful characters but as resources that constitute social interaction~\cite{boellstorff2015secondlife}.
In social VR, multiple users communicate in real-time through embodied avatars~\cite{freeman2021socialvravatar}, and strong co-presence can be achieved via bodily cues, such as voice and gesture~\cite{maloney2020socialvrnonverbalcscw,tonenbaum2020socialnonverbalchi}.
Consequently, avatar design and platform-specific affordances collectively influence what forms of communication and expression are possible~\cite{baker2021amc,kim2023amc,nowak2018amc}.

\added{Avatar-mediated performance has become increasingly prominent, with VTubing emerging as one of its most recognizable forms. VTubers are livestreamers who create content through animated 2D or 3D avatars, allowing them to perform without exposing their physical bodies. This disembodiment affords substantial flexibility in identity construction and performance style, enabling exaggerated self-presentation and rich audience interaction~\cite{wan2024vtuber}. VTubing also extends research on self-presentation and aligns with the Proteus effect, which posits that avatar characteristics can shape users' attitudes and behaviors~\cite{yee2007proteus}.}

\added{While VTubing relies on 2D/3D animated avatars, social VR introduces fully embodied avatars that convey nonverbal cues such as speech, gesture, and spatial positioning, supporting heightened co-presence and more immediate, bidirectional interaction~\cite{baker2021amc}. In social VR, performers may receive stronger signals of social presence from their audiences, as audiences are embodied as avatars. This potentially allows for a more direct assessment of real-time audience reactions compared to traditional VTubing environments.}

\added{However, despite this promise, behavioral human-likeness in social VR remains limited. Nonverbal channels such as lip movement and gaze direction often depend on external plugins and are not yet reliably mapped to users' real behaviors~\cite{tonenbaum2020socialnonverbalchi,freeman2021socialvravatar,baker2021amc}. As a result, it is still unclear how performers interpret audience feedback or assess engagement when both parties are represented through embodied, but imperfectly expressive, avatars.
Thus, this study aims to examine how performers interact with and interpret their remote audiences when both sides are mediated through embodied avatars in social VR.}

\subsection{\added{Technology-Mediated Comedy, Laughter, and Norms}}
Comedy is a live performance genre that relies on a rich exchange of social cues between performers and their audiences.
Stand-up comedy is typically a solo act that addresses the audience with minimal reliance on costumes or props~\cite{mintz1985standup}.
Conversation-analytic work shows that even ostensibly verbal performances rely on nonverbal and paralinguistic signals to project laughability and synchronize audience responses~\cite{scarpetta2009interactionalcomedy}.
Techniques described as ``laugh-traps''~\cite{mcilvenny1992laughtraps} are enacted through gesture~\cite{katevas2015robotcomedy}, eyebrow movements~\cite{tabacaru2014eyebrows}, facial expressions~\cite{tabacaru2014facialhumor}, shifts in voice quality and prosody~\cite{rutter1997standup,rutter2001rhetoric}, and gaze~\cite{scarpetta2009interactionalcomedy}.
These practices parallel patterns in ordinary conversation where speakers use multimodal cues such as gaze, gesture, and prosody to invite recipients to respond~\cite{goodwin1979naturalconversation,kendon1967socialinteraction}.
When laughter is sparse, comedians adjust material and pacing in situ~\cite{wells2007audiencereaction,harbidge2011audienceship}.
Comparable cue-based coordination is also observed in two-person formats, such as British double acts~\cite{hewett2021doubleact} and Japan's \textit{manzai}~\cite{katayama2006manzai,katayama2008manzai}.
Beyond the role of shared background knowledge emphasized in humor theory~\cite{glick2007standupcomedytechniques}, reception is shaped by local, in-the-moment contexts~\cite{scarpetta2009interactionalcomedy}.

\added{Prior research has explored interventions that strengthen laughter contagion and enhance the visibility of audience feedback when co-presence is limited. For example, laugh tracks demonstrate how inserted or recorded audience laughter can cue humor and influence audience perception~\cite{giotta2017laughtrack}. Building on this, computational approaches have developed adaptive laugh tracks that trigger playback contingent on detected viewer laughter~\cite{fukushima2010laughtrack} or that incorporate nonverbal cues beyond audio to convey a richer sense of sociality while watching comedy~\cite{ohara2025laughtrack}. Related work also examines how avatar-expressed laughter can transform solitary media consumption into more socially resonant viewing experiences, even when audiences are physically distributed~\cite{ohara2025cscw}.}

\added{Social VR platforms enable users to interact with performers through fully embodied avatars, creating new opportunities for participatory comedy and collective enjoyment. However, this capability also introduces vulnerabilities. Several studies have documented issues such as misconduct \cite{chen2025moderation}, harassment \cite{freenan2022harassment,schulenverg2023harassment}, and boundary violations \cite{blackwell2019harassment} within social VR. These include problems like intrusive proximity, unwanted touching via avatars, mimicked gestures, verbal harassment, and disruptive audience behaviors that can compromise users' sense of safety and agency \cite{maloney2021socialvrethic}. Researchers have found that the lack of clarity around social norms, the anonymity provided by avatars, and the challenges of moderating embodied interactions contribute to an increased risk of harassment and impersonation in social VR \cite{blackwell2019harassment}. These challenges become even more pronounced in performance settings, where performers must continually track audience engagement while maintaining the quality of their performances.}

\added{Yet, despite growing concerns about misbehavior and safety \cite{blackwell2019harassment,freenan2022harassment,maloney2021socialvrethic}, little is known about how specific creative communities, such as comedy performers, cultivate social norms, regulate audience behavior, and co-create shared expectations in social VR. Comedy is a particularly revealing domain because it is inherently interactional, often improvisational, and highly dependent on timing and audience feedback. Comedic performance also frequently invites playful transgression, making it an ideal context for observing how communities establish norms to regulate acceptable behavior and misconduct.}



\section{METHODOLOGY}
\subsection{Positionality Statement}
Positionality refers to researchers' acknowledgment of how their social and professional backgrounds shape the research process~\cite{bourke2014positionality,singh2025positionality}.
In social VR research, dynamics of embodiment and anonymity 
\added{shape how researchers are perceived and what interactions they can access, highlighting the need for situated engagement~\cite{maloney2021socialvrethic}.}
In this study, positionality was salient because the first author entered the field as an outsider to and gradually became a partial insider through nine months of involvement, including performing on stage.

The first author, who conducted all data collection and led the primary analysis, had no prior experience with social VR at the outset of the study. 
To gain contextual understanding, the author engaged with a VRChat comedy community, 
\added{and these experiences facilitated trust building and provided situated insights for interpreting participants' accounts.}
The co-authors provided complementary insider and outsider perspectives,
 as one CSCW researcher with no social VR experience questioned taken-for-granted assumptions and another with long-term VRChat experience assessed whether the plausiblity of cultural interpretations.
\added{Two additional co-authors with expertise in social VR contributed in ongoing discussions of the study design and emerging themes, supporting a reflexive analytic process.}

\subsection{Participants and Study Sites}
\begin{table*}[htbp]
  \centering
  \captionsetup{width=0.95\textwidth}
  \caption{Demographic information of interview and observation participants. ``--'' indicates no response from interview participants, and ``†'' indicates that the information was not available due to observation-only participation.}
  \label{tab:participant_info}
  \begin{tabular}{lcccccccc}
    \hline
    ID &
    \begin{tabular}[c]{@{}c@{}}Evaluation\\Method\end{tabular} &
    \begin{tabular}[c]{@{}c@{}}Main Perf.\\Type\end{tabular} &
    \begin{tabular}[c]{@{}c@{}}Avatar\\Type\end{tabular} &
    Platform &
    Gender &
    Age &
    \begin{tabular}[c]{@{}c@{}}Play\\Time\end{tabular} &
    \begin{tabular}[c]{@{}c@{}}Real-world\\Experience\end{tabular} \\
    \hline\hline
    P01 & Interviewed \& Observed & Stand-up & Humanoid & VRChat, Cluster & Male & 35 & 15000 & None \\
    P02 & Interviewed \& Observed & Solo-sketch & Humanoid & VRChat, Cluster & Male & -- & 2000 & Several times \\
    P03 & Interviewed \& Observed & Manzai & Humanoid & VRChat & Male & 38 & 2833 & Occasionally \\
    P04 & Interviewed \& Observed & Rakugo & Semi-Humanoid & VRChat, Cluster & Male & -- & 3500 & None \\
    P05 & Interviewed \& Observed & Solo-sketch & Semi-Humanoid & VRChat, Cluster & Male & 32 & 5000 & None \\
    P06 & Interviewed \& Observed & Solo-sketch & Humanoid & VRChat, Cluster & Female & -- & 279 & None \\
    P07 & Interviewed \& Observed & Manzai & Semi-Humanoid & VRChat, Cluster & Female & 34 & 800 & Professional \\
    P08 & Interviewed \& Observed & Rakugo & Semi-Humanoid & VRChat, Cluster & Male & 25 & 1000 & Professional \\
    P09 & Interviewed \& Observed & Manzai & Humanoid & VRChat & Female & 49 & 2000 & Professional \\
    P10 & Interviewed \& Observed & Manzai & Humanoid & VRChat & Male & 24 & 500 & None \\
    P11 & Interviewed \& Observed & Manzai & Humanoid & VRChat, Cluster & Male & 20s & 1900 & Occasionally \\
    P12 & Interviewed \& Observed & Manzai & Semi-Humanoid & VRChat, Cluster & Male & 28 & 1977 & Occasionally \\
    P13 & Interviewed \& Observed & Rakugo & Humanoid & VRChat, Cluster & Male & 30 & 9000 & None \\
    P14 & Interviewed \& Observed & Solo-sketch & Humanoid & VRChat, Cluster & Male & 54 & 1200 & None \\
    P15 & Interviewed \& Observed & Stand-up & Humanoid & VRChat, Cluster & Female & -- & 944 & Occasionally \\
    P16 & Observed & Solo-sketch & Semi-Humanoid & VRChat & † & † & † & † \\
    P17 & Observed & Stand-up & Humanoid & VRChat & † & † & † & † \\
    P18 & Observed & Stand-up & Semi-Humanoid & VRChat & † & † & † & † \\
    P19 & Observed & Manzai & Humanoid & VRChat & † & † & † & † \\
    P20 & Observed & Manzai & Humanoid & VRChat & † & † & † & † \\
    P21 & Observed & Manzai & Humanoid & VRChat & † & † & † & † \\
    P22 & Observed & Stand-up & Semi-Humanoid & VRChat & † & † & † & † \\
    P23 & Observed & Manzai & Humanoid & VRChat & † & † & † & † \\
    \hline
  \end{tabular}
\end{table*}

This study examined virtual comedians' performance practivces and audience interactions through semi-structured interviews and performance observations.
Interviews and observations were conducted with 15 virtual comedians (P01–P15). 
To triangulate the findings, we also analyzed videos of performances from eight performers (P16–P23), totaling 23 participants.
Details of participant demographics and performance attributes are shown in Table~\ref{tab:participant_info}.

Participants were recruited from ``VRC Owarai Dojo\footnote{VRC Owarai Dojo YouTube Channel \url{https://www.youtube.com/@vrc8765}},'' a virtual comedian community in VRChat in Japan.
All participants had experience in performing comedy on VRChat, and twelve also performed on Cluster.
\added{Although recruited from a single community, participants regularly performed at independent events and across multiple social VR settings, indicating that their practices extended beyond the norms of one group.}
Three participants were also professional comedians in the physical world.
Among the fifteen interview participants, eleven identified as male and four as female.  
This gender distribution roughly reflects that of the general metaverse user population~\cite{nemu2023socialvr}.
\added{The performances observed in this study included four comedy styles: stand-up, solo-sketch, manzai, and rakugo, with brief explanations and illustrations provided in Appendix~\ref{appendix:comedystyle}.}
The experimental protocol was approved by the local ethics committee of the authors' institution.

\added{Stand-up involved a single comedian delivering a spoken monologue with a microphone.}
\added{Solo-sketch featured a single performer enacting a short narrative, sometimes using props or visual effects.}
Manzai was performed by two comedians engaging in a rapid back-and-forth dialogue, usually structured around the roles of ``straight person'' and ``funny person''~\cite{katayama2006manzai,katayama2008manzai}.
Rakugo, a traditional form of seated storytelling, involved a single performer who portrayed multiple characters within a scripted story~\cite{sakai1981rakugo}.

Participants' primary avatar types were classified according to 
\added{a large-scale survey of social VR users}~\cite{nemu2023socialvr}, which reports several avatar style categories, among which only Humanoid and Semi-Humanoid types were represented in our participant sample.
In the observed events, individual performances lasted between three to ten minutes, with five to six comedians per show.
Audience size ranged from 50 to 80 people, which aligns with the maximum capacity of one VRChat instance\footnote{As of September 2025, according to the VRChat release notes: \url{https://hello.vrchat.com/blog/april-dev-update}.}.
Performances took place in stage-like VRChat worlds.

\subsubsection{Interviews}
Interview participants were recruited using purposive sampling~\cite{robinson2014sampling}.
We initially invited participants who had experience in social VR through the community's Discord server and email channels.
Interviews were conducted with the first five individuals who responded.
\added{An initial analysis indicated the need for diversity in physical-world performance experience and gender, as professional comedians offered comparisons between live and social VR performance and gender shaped motivations for avatar selection. We therefore directly invited additional participants.}
Data were analyzed iteratively, and recruitment concluded when theoretical saturation was reached~\cite{braun2021saturate}.

Interviews were conducted in private VRChat worlds\footnote{``World'' refers to a virtual room in VRChat} accessible only to the researcher and the participant.
Participants received an interview overview in advance and provided explicit consent by signing the informed consent form and agreeing to video and audio recording.
During the interview, the researcher used their usual performer avatar, and participants could use any avatar they preferred.
Participants occasionally switched avatars or demonstrated performance-related gimmicks.
Each interview lasted 60–90 minutes.
Interviews were recorded using OBS Studio\footnote{OBS Studio \url{https://obsproject.com/}}, capturing both audio and video.  
Participants received a 3,000 JPY Amazon gift card as compensation.
The complete interview guide is provided in Appendix~\ref{appendix:interviewguide}.

\subsubsection{Observations}
To triangulate the interview data, we analyzed publicly available recordings of the comedians' performances. 
We collected YouTube recordings for all 15 interview perticipants.
To reduce sampling bias, we also included eight additional virtual comedians who had at least three publicly available recordings.
For each comedian, we analyzed three recordings representing early, middle, and recent stages of their performance history.
Early recordings often included debut or introductory moments that showed initial adaptation to social VR.
These recordings provided a baseline for understanding subsequent performance development.
When we observed notable shifts in performance style, we analyzed additional recordings for that performer.

In total, we analyzed 71 recordings (mean duration 7.46 minutes).
\added{For each recording, the first author created time-stamped analytic memos documenting performer actions, audience reactions, and key visual elements.}
\added{We treated each recording as a coherent unit of analysis and iteratively coded segments relevant to our research questions. Similar video-based qualitative approaches have been used in studies of VTuber performance~\cite{wan2024vtuber}.}
This process allowed us to examine narrative structures, reaction timing, and visual affordances in context.

Guided by RQ1 and RQ2 and by repeated viewing of the performance videos, we focused on four aspects:
\begin{itemize}
    \item How avatar design, bodily movement, and virtual environments or props shaped comedic performance.
    \item How audience reactions were expressed and how performers adjusted to them.
    \item How timing affected performance and interaction including issues between performer and audience.
    \item How instances of disruption occurred, such as technical glitches or social interruptions.
\end{itemize}
These focal points were also informed by interview findings highlighting the importance of temporal dynamics in social VR performance. 
For each video, we compiled transcriptions, timestamps, and screenshots as supplementary materials for analysis.

According to the coding scheme, we selected three recordings for each virtual comedian.
These recordings represented early, middle, and recent stages of their performance history.
Early performances often included introductory moments noted as important in VTuber research~\cite{wan2024vtuber}.
Recent performances reflected accumulated experience and adaptation to social VR.
When performance strategies differed markedly across these stages, we analyzed additional recordings to capture these changes.

\subsection{Data Analysis}
The qualitative data consisted of interview transcripts and observation notes derived from video recordings.  
We analyzed the data using reflexive thematic analysis~\cite{braun2019thematic}, \added{adopting a primarily deductive orientation informed by our research questions and allowing inductive codes to emerge when novel patterns appeared.}  
First, the first author transcribed the interview data verbatim.  
The first author then repeatedly read the transcripts and observation notes to gain familiarity with the dataset and generated initial codes.  
The second and third authors joined the analysis to review codes and discuss their relationships, during which codes were revised and themes were iteratively refined.  
\added{Throughout the analysis, the authors engaged in reflexive discussion about how their positionalities influenced interpretive decisions, consistent with the principles of reflexive thematic analysis.}  
Themes were finalized through consensus among all authors.  
\added{Coding continued until no new patterns relevant to the research questions emerged.}  
The analysis was conducted using ATLAS.ti\footnote{ATLAS.ti \url{https://atlasti.com/}}.  
All materials (transcripts, notes, codes, and themes) were translated into English for reporting in this paper.

\section{FINDINGS}
As a result of our thematic analysis, we identified seven main themes and fourteen sub-themes.
These are summarized in Table~\ref{tab:theme_summary}. 

\begin{table*}[htbp]
  \centering
  \captionsetup{width=0.95\textwidth}
  \caption{Research Questions and Identified Themes}
  \label{tab:theme_summary}
  \begin{tabular}{lll}
    \toprule
    \multicolumn{1}{c}{\textbf{RQ}} & \multicolumn{1}{c}{\textbf{Theme}} & \multicolumn{1}{c}{\textbf{Subtheme}} \\
    \midrule
    \multirow{6}{*}{\begin{tabular}[c]{@{}l@{}}Performers' \\ Strategies (RQ1)\end{tabular}}
      & Creative Transformation of Constraints & Manually Expressing Nonverbal Cues \\
      &                                        & Exaggerating Performative Expressions \\      
      \cmidrule(lr){2-3}
      & Strategically Adjusting Avatars for Performance & Avatar Use as Social Affordance for Joint Performance  \\
      &                         & Avatar Switching for Psychological and Character Shifts \\
      \cmidrule(lr){2-3}
      & Dependence on Technical Expertise & Lack of Technical Skills \\ 
      &                                        & Reappropriating Technical Glitches to Improvisational Humor \\      
    \midrule
    \multirow{4}{*}{\begin{tabular}[c]{@{}l@{}}Interaction (RQ2)\end{tabular}}
      & Audience Feedback under Expressive Constraints & Absence of Nonverbal Feedback \\
      &                            & Context-appropriate Emoji Reactions \\
      \cmidrule(lr){2-3}
      & Disrupted Timing of Audience Feedback & Latency in Vocal Reactions \\
      &                                     & Unreliable Timing of Intentional Reactions \\
    \midrule
    \multirow{4}{*}{\begin{tabular}[c]{@{}l@{}}Community \\ Practices (RQ3)\end{tabular}}
      & Community-based Moderation Practices & Managing Disruptions and Trolling \\
      &                      & Community Self-Governance \\
      \cmidrule(lr){2-3}
      & VR as Shared Epistemic Ground & Using VR as Material for Jokes \\
      &                                & Challenges of Community Expansion \\
    \bottomrule
  \end{tabular}
\end{table*}

\subsection{Strategies for the Performance of Virtual Comedians}
\subsubsection{Creative Transformation of Constraints}
\label{subsubsec:4-1-1}
Performing comedy in social VR involves expressive limitations that differ fundamentally from those in physical stage environments. 
Rather than treating these constraints as deficiencies, virtual comedians incorporated them into creative opportunities. 
Some performers manually triggered emotional expressions or exaggerated avatar movements, allowing them to generate laughter through representational logics unique to virtual environments.
We elaborate on each point below.

\paragraph{Manually Expressing Nonverbal Cues}
Many virtual comedians acknowledged that their avatars' expressive capabilities were more limited than those of the physical body, yet they enacted expressive performances by manually controlling nonverbal cues. 
P13, a performer specializing in rakugo, explained:

\begin{quote}
\textit{``In rakugo, a performer has to play two characters alone, using only posture and facial expression. When switching roles quickly, expressions alone are often used to distinguish the characters. But in VR, it's difficult to convey expressions--I think I am managing relatively well, though (laughs). I use hand signs to toggle between pre-defined facial expressions [that I created before performing in VR]. For traditional rakugo performers, this might look strange. They might wonder why I have to move my fingers when the moment clearly calls for a facial shift.''}
\end{quote}

In our observation of P14, we found similar techniques of mapping manual gestures to facial expressions. 
They configured their avatar so that spreading their hands outward would trigger an automatic transition to a smiling facial expression (Figure~\ref{fig:teaser} (B)). 
This mapping allowed them to control facial affect with minimal cognitive load, enabling them to maintain rhythm and timing during live solo acts. 

Some comedians pointed out that their ability to compensate for limited nonverbal cues depended on the affordances of different social VR platforms. 
P02 noted:

\begin{quote}
\textit{``In the real world, you can quickly shift expressions to evoke laughter, but I can't do that here [in VRChat]. So I fix my avatar's face to a neutral expression and deliberately change it using the controller depending on the scene. When I want to perform seriously and control my avatar's face with precision, VRChat is the best. But with Cluster, I can use emotes from a smartphone or load images natively, so there are performances that can only happen there.''}
\end{quote}

These findings revealed that virtual comedians actively leverage the affordances of social VR platforms to compensate for the limited expressiveness of avatars and achieve performances that would not be possible in the physical world.

\paragraph{Exaggerating Performative Expressions}
Some virtual comedians leveraged the limited resolution of bodily movements in social VR as a resource for exaggeration.
Instead of reproducing fine-grained gestures as in face-to-face settings, they amplified simple actions into large, stylized movements that were more easily perceived by the audience.
P09, a professional physical comedian, described how movement tracking in social VR required them to adapt familiar routines through exaggerated actions:

\begin{quote}
\textit{"What I've realized from performing in both real and virtual spaces is that many expressions I use in physical comedy simply don't carry over in VR. So I change how I perform. For instance, when I do tsukkomi (the straight man's punchline in manzai comedy), I make them stronger, sometimes my partner even flies away. That creates a clear impression for the audience. But this requires full-body tracking, not just a headset and hand controllers."}
\end{quote}

We also observed that virtual comedians (e.g., P12, P13) incorporated exaggerated, physically implausible movements, such as floating or bouncing, to simulate conditions like zero gravity (Figure~\ref{fig:teaser} (A)).
Based on body posture and movement trajectories, these effects appeared to result from improvisational physical manipulation in the real world, such as reclining in a chair to create the illusion of floating.

Interestingly, even genres like rakugo, which traditionally emphasize subtle gestures, were adapted using exaggerated expressions. P02, typically a solo-sketch performer, reflected on their experience adapting rakugo to VR:

\begin{quote}
\textit{"I've never done rakugo in the real world—only in social VR. But I always try to exaggerate my gestures. Since I don't use full-body tracking, trying to replicate nuanced movements would only highlight my limitations. So instead, I use broad, cartoon-like movements that work better in VR."}
\end{quote}

These examples illustrated that virtual comedians were not merely compensating for the limitations of avatar-mediated performances, but instead developing a distinct performative vocabulary for social VR. 

\subsubsection{Strategically Adjusting Avatars for Performance}
\label{subsubsec:4-1-2}
Avatars serve as a central medium of self-presentation in digital environments, enabling users to convey aspects of their identity, personality, and social affiliation through customizable visual features.
In contrast, virtual comedians often prioritize the functional role of avatars in performance contexts, strategically modulating visual features to avoid unnecessary computational load and distractions from the act.
For example, virtual comedians adjusted their avatars to align with those of their co-performers or toned down aesthetic embellishments to maintain visual coherence on stage.
This flexible modulation enabled them to maintain a consistent performer identity while flexibly adapting to diverse content, formats, and collaborative arrangements.

\paragraph{Avatar Use as Social Affordance for Joint Performance}
Virtual comedians tended to prefer humanoid avatars with minimal ornamental features.  
They generally avoided exaggerated appearances or non-human designs.
In our observations, most comedians (e.g., P17, P19) used plain avatars with no flashy attributes. They did not use visual features to build a unique character identity or generate humor through appearance.
Instead, the comedic effect was primarily achieved through narrative structure, verbal delivery, and timing.

This visual neutrality was not simply based on aesthetic preference.
Performers appeared to utilize it as a strategy to enhance expressive flexibility and maintain visual coherence with co-performers across various settings and genres.
Comedians who frequently performed in groups, such as P07, paid particular attention to visual coherence among collaborators.

\begin{quote}
\textit{``To make sure we look visually coherent, I sometimes switch to an avatar with the same design style or from the same creator as my partner. When performing with a male comedian, I may even switch to a male avatar.''}
\end{quote}

These practices suggest that avatar appearance in social VR comedy functions as a social affordance that shapes both how performers are perceived by audiences and how they coordinate with collaborators.
Through intentional visual adjustments, performers supported shared attention, facilitated interactional alignment, and enhanced the overall coherence of distributed performances.

\paragraph{Avatar Switching for Psychological Shifts}

Several virtual comedians reported that they regularly switched between avatars used in everyday social VR settings and those used specifically for performance.
These shifts were not merely cosmetic but served as a psychological trigger to enter a performative mindset.
P15 noted the motivational role of avatar changes. They explained that specific outfits were reserved for comedy performances:

\begin{quote}
\textit{``When I'm not performing in VRChat, I don't wear the outfit. The oiran (a Japanese courtesan-style) costume is only for comedy. I want to act like a regular user in everyday settings. When I perform, wearing the costume helps me feel bright and energetic. It helps me both mentally and commercially.''}
\end{quote}

P02 also emphasized the effect of costume changes as follows:
\begin{quote}
\textit{``When I change my avatar, I go into comedian mode. It happens the moment I switch. My mood is pulled by the avatar, and it helps me shift my mindset.''}
\end{quote}

These reflections resonate with the Proteus effect, which refers to the phenomenon where an individual's behavior tends to conform to the traits or social expectations associated with their avatar~\cite{yee2007proteus}.
This suggests that the Proteus effect may also operate in live comedy performance settings within social VR, where switching avatars serves not only as a visual transformation but also as a psychological mechanism for enacting distinct comedic personas.

\subsubsection{Dependence on Technical Expertise}
Our findings revealed that disparities in technical proficiency directly affected the kinds of performances comedians were able to execute.
Some participants expressed frustration over the limitations imposed by their lack of technical skills. Others, who were more technically proficient, were able to incorporate system glitches into improvisational humor. 

\label{subsubsec:4-1-3}
\paragraph{Lack of Technical Skills}
Several virtual comedians noted that a lack of technical expertise constrained the range of performances they could deliver in social VR.  
While the medium offers unique affordances for expressive staging, realizing these often requires familiarity with tools such as Unity or VRChat's scripting systems.
P09, a professional comedian in the physical world, reflected on such limitations:

\begin{quote}
\textit{``I want to do performances that are only possible in social VR, but I don't have the technical skills. There was a moment where I wanted to use particles during a performance, but I had no idea how to implement them.''}
\end{quote}

Due to these technical constraints, P09's performances consisted solely of verbal routines.
Without the ability to implement visual effects or motion-based gags, the comedic structure relied entirely on speech, limiting the expressive possibilities unique to the medium.

These accounts suggest that access to technical skills has a significant influence on the creative freedom of performers.  
Those with prior experience or a willingness to learn development tools could expand their expressive repertoire, while others remained limited by the difficulty of implementation.

\paragraph{Reappropriating Technical Glitches to Improvisational Humor}
In social VR performances, technical disruptions such as disconnections from shared spaces due to network congestion were frequently reported. Rather than treating these as performance failures, virtual comedians often framed such glitches as opportunities for improvisational humor. 

P03, who performs as part of a comedy duo, recounted how their co-performer improvised when technical issues occurred in VR:
\begin{quote}
\textit{``While switching avatars during our performance, the bandwidth spiked and I got kicked out of the world (the virtual space in VRChat). While I couldn't continue, my partner kept the audience entertained with improvised banter.''}  
\end{quote}

Similarly, P01 reflected that:  
\begin{quote}
\textit{``The comedians who become popular here are the ones who can immediately turn tech troubles into jokes on the fly.''}  
\end{quote}
While improvising around audience reactions or venue events is not uncommon in live comedy more broadly, what distinguishes social VR is the heightened need to adapt to platform-specific technical issues. The ability to incorporate such disruptions into the act emerges as a unique competency for virtual comedians, underscoring how system fragility becomes a performative resource.

\subsection{Interactions Between Virtual Comedians and Their Audiences}
\subsubsection{Audience Feedback under Expressive Constraints}
\label{subsubsec:4-2-1}
In physical comedy, performers rely on a variety of audience cues, such as facial expressions, gestures, and body movements, to adjust their delivery in real-time. 
By contrast, in social VR, these nonverbal cues are limited, which makes it difficult for comedians to gauge the audience's reactions in real-time. 
Our findings revealed that nonverbal cues are severely limited in social VR, which often left performers uncertain about the level of engagement because subtle laughter or gestures could not be reliably perceived, and that audiences instead developed distinctive practice of context-appropriate emoji reactions as a new form of feedback.

\paragraph{Absence of Nonverbal Feedback}
Many virtual comedians pointed out that they lacked sufficient nonverbal cues to obtain feedback from the audience during their performances.  
P03, who also works as a professional comedian, stated,  
\begin{quote}
\textit{``When I am performing, it is hard to tell whether the audience is really laughing. [When they use an HMD] in VR mode, I can somewhat see if they lean back while laughing, but there are clear limitations. Even then, their facial expressions do not change.''}
\end{quote}
On many social VR platforms, audiences can also participate from PCs or smartphones without using HMDs.  
As P03 described, even when audience members use HMDs nonverbal cues are still limited, and when they do not use them such cues are virtually absent, creating a significant constraint for performances in social VR.  

\paragraph{Context-appropriate Emoji Reactions}
As a compensatory form of interaction for the lack of nonverbal cues in social VR, we observed audience feedback through the use of emoji.  
For example, in VRChat, users can open a menu (Figure~\ref{fig:emojis}A) and tilt the joystick to select from 66 predefined emoji.  
This function was frequently used in comedy performances as a means of sharing emotions.  

What was particularly notable was not only the use of laughter emoji at humorous moments, but also the distinctive practice of sending emoji that matched the content of the performance.  
In Figure~\ref{fig:emojis}B, for instance, the comedian performed a solo sketch in which they dropped a wallet.  
At that moment, as shown in (Figure~\ref{fig:teaser} (C)), audience members responded not only with laughter emoji but also with money emoji.  
In this way, despite the challenges of text input and the absence of spontaneous nonverbal cues, audiences developed distinctive interactional practices specific to social VR.  

\subsubsection{Disrupted Timing of Audience Feedback}
\label{subsubsec:4-2-2}
Virtual comedians emphasized that one of the key challenges of performing in social VR was the disruption in the timing of audience feedback.  
This issue can be broadly divided into two types: delays in laughter, which is a natural and spontaneous form of feedback, and delays in intentional feedback such as emoji.  

\paragraph{Latency in Vocal Reactions}
In live comedy, performers often adjust their delivery and improvisation based on the immediate reactions of the audience.  
However, in social VR such techniques were frequently reported to be unworkable due to audio latency.  
P09, a professional comedian in the physical world, explained,  
\begin{quote}
\textit{``In social VR, audio is delayed, so I cannot use the technique of waiting for the audience to finish laughing before moving on to the next part of the performance. If I wait for the laughter to end, I feel the audience will lose interest.''}  
\end{quote}

Similarly, P08, also a professional comedian, highlighted the same issue,  
\begin{quote}
\textit{``In comedy performance, audience laughter is essential. With software like Syncroom, it is possible to reduce audio latency among performers, but the delay with the audience cannot be solved. That is why I adapt my performance so that it does not rely on waiting for laughter. Instead of delivering punchlines sharply, I try to say funny things more slowly.''}  
\end{quote}

These accounts show that even professional comedians accustomed to precise timing in live venues had to abandon established strategies such as ``waiting for laughter'' and instead invent alternative pacing styles that fit the delayed environment of social VR.

\paragraph{Unreliable of Intentional Reactions}
While intentional forms of feedback such as emoji were distinctive to social VR, comedians pointed out that such feedback raised concerns about timing and whether it truly reflected the audience's feelings. 
P03, comparing the platforms VRChat and Cluster, explained,  
\begin{quote}
\textit{``In Cluster, many audience members join from smartphones, so most of them keep their microphones muted. As a result, feedback mainly comes through emoji or visual effects. But I sometimes feel that these are pressed intentionally rather than as a natural response. It depends on the person, but I wonder if they are pressing the button even when they are not actually laughing. When you laugh spontaneously, do you really take the time to press a button?''}  
\end{quote}

This highlights a broader uncertainty around the authenticity of intentional reactions.
Performers sometimes questioned whether audiences were genuinely amused or merely signaling through interface actions, which undermined the spontaneity that usually anchors performer–audience interaction in live comedy.

In this way, both the temporal delay of natural laughter and the questionable reliability of intentional feedback emerged as key challenges in how comedians experienced audience responses in social VR, directly affecting their ability to manage rhythm, timing, and authenticity on stage.

\subsection{Shared Community Practices in Social VR Comedy}
\label{sec:rq3}
\subsubsection{Community-based Moderation}
\label{subsubsec:4-3-1}
Unlike physical stages, performances in social VR were often reported to be interrupted by unexpected disruptions or trolling.  
In response, our findings revealed that community-based moderation was carried out by virtual comedians and regular audience members.  

\paragraph{Managing Disruptions and Trolling}
Performers frequently mentioned cases where audience behavior, including disruptive actions and trolling, such as speaking loudly during the performance, interfered with the progress of performances.  
P01 reflected on situations in social VR where lengthy spoken comments from the audience disrupted the flow of the performance: 
\begin{quote}
\textit{``There were audience members who gave long comments or spoke too loudly during the act. Once, someone shouted, ``That could never happen!'' in response to a joke. Depending on the content, that kind of reaction can make comedy more interesting. But sometimes, it just gets in the way of the performance.''}
\end{quote}

P06 pointed out that such disruptive audience members seemed more common compared to physical stages:  
\begin{quote}
\textit{``Compared to live comedy shows in the physical world, I feel there are more people who speak out or interrupt during performances. I think the reason is that in online settings, anonymity makes it easier to speak up.''}
\end{quote}

Such inappropriate audience conduct emerged as a distinctive challenge for virtual comedy performances. 

\paragraph{Self-Governance by Communities}
Unlike physical comedy venues, where staff or security personnel are present, no such formal oversight exists when performing in social VR.  
In this context, governance practices to handle trolling were established collectively by virtual comedians and regular audience members. They would share information within online communities to identify individuals who frequently engage in misconduct.
For example, P06 explained about users who had previously disrupted performances intentionally,
\begin{quote}
\textit{``Information about users who troll is often shared through community Discord servers or word-of-mouth among frequent social VR users. When such users appear during a performance, comedians and audiences cooperate by using VRChat's kick function to deal with them.''}
\end{quote}

VRChat allows users to randomly enter publicly accessible worlds, which often causes unexpected encounters and makes it difficult for virtual comedians to concentrate on their performance.
Reflecting on this challenge, P05 explained:

\begin{quote}
\textit{``When we are performing or recording an archive, we cannot always deal with troublemakers ourselves. To prevent disruptions, we sometimes set the instance to restricted modes, such as Friend+ or Group+.''}
\end{quote}

This practice illustrates how virtual comedians collectively leverage access-restricted instances, such as Friend+ or Group+\footnote{In VRChat, a Friend+ instance is accessible to the host's friends and their friends, while a Group+ instance is limited to members of a designated community group.}, to reduce interruptions and sustain a stable performance environment.
 

\subsubsection{VR as Shared Epistemic Ground for Comedy}
\label{subsubsec:4-3-2}
Our findings revealed that VR itself functioned as a shared epistemic ground that connected comedians and audiences.  
Many virtual comedians deliberately incorporated VR or social VR as the subject matter of their performances, drawing on experiences and references that audiences were expected to share.  
This mutual grounding supported quick recognition of humor but also raised challenges when the community expanded to include newcomers unfamiliar with these references.

\paragraph{Using VR as Material for Jokes}
Comedy often relies on shared cultural knowledge as the foundation for humor, and in social VR, the environment itself frequently provides this common ground.
For example, traditional Japanese comedy such as \textit{rakugo} is based on well-known scripts that have been performed for generations, with performers following established storylines on stage.  
P04, who performed rakugo in social VR, explained how they adapted one such classic story, \textit{Shibahama}, which normally involves a character picking up a wallet full of money.  
\begin{quote}
\textit{``There is a famous rakugo script called ``Shibahama.'' When I perform it in VR, the flow follows the script, but instead of picking up money, I pick up a Meta Quest headset. Everyone in VRChat understands this, so they laugh. Because it is a social VR community, the reference makes sense to all of them.''}
\end{quote}
In another performance by P04, the comedian used VRChat's radial menu (the interface for selecting avatar expressions) as a source of humor.  
They joked about how the menu had so many options that the radial divisions became comically thin, a situation immediately recognizable to anyone who regularly uses the platform.  
This use of VR as a shared cultural resource enabled comedians to secure quick audience recognition and collective laughter, illustrating how the medium itself shaped the content and reception of humor in social VR.

\paragraph{Challenges of Community Expansion}
Participants reported that social VR comedy communities have rapidly grown in recent years, but this growth has also introduced new challenges related to technical constraints and the influx of audiences who do not share the same cultural premises as long-term participants.
VRChat restricts the number of simultaneous users who can enter a given virtual world (an ``instance''), and once this limit is reached, additional users are unable to join until others leave. 
P13 described a direct consequence of VRChat's instance capacity limits, which are capped at 80 users:  
\begin{quote}
\textit{``It reached the maximum instance size, and even though I was scheduled to perform, I couldn't get in. In the end my turn was skipped. I only managed to join once the audience numbers dropped a little.''}
\end{quote}
As P13's account illustrates, such technical restrictions sometimes prevented both performers and audiences from participating as intended.  

Concerns about scaling were also raised regarding community management.  
P04 explained that the small and relatively closed nature of VRChat communities currently facilitates shared understanding:  
\begin{quote}
\textit{``Right now, VRChat itself is a limited, closed community, so it's always easy to align topics. If you use VR as an example, everyone immediately understands. [...] But as the audience grows, organizers will also be needed. At the moment, there are only comedy performers, and no one is taking care of [community] management.''}
\end{quote}
This highlights how community expansion not only risks weakening the shared cultural background that helps participants understand jokes, but also increases the demand for organizational roles that extend beyond performing itself.

\section{DISCUSSION}
\subsection{Making Constraints into Opportunities in VR Performance (RQ1)}
Our findings showed that virtual comedians often transformed the constraints of avatar-mediated environments into resources for their performances (Section~\ref{subsubsec:4-1-1}, \ref{subsubsec:4-1-2}). Instead of being hindered by technical limitations, they creatively reframed them as opportunities. For example, some performers deliberately exploited the lack of real-time facial tracking. They maintained a neutral expression and then manually switched to exaggerated expressions of anger or surprise to generate a comedic effect. Others experimented with physical staging in their real-world environment: by lying on a chair, an object not captured in VR, they could simulate the illusion of floating in space, producing humor through impossible gestures. 
Collectively, such practices highlight how performers actively reinterpret the affordances within social VR, distinguishing themselves from comedians in physical settings who cannot rely on similar manipulations.

This creative reframing resonates with the notion of "creative constraints" proposed in the context of online theatre by \citet{lc2023contradiction}. \citeauthor{lc2023contradiction} observed that online theatre performers reinterpreted their inability to see audience reactions as a positive condition that reduced performers' cognitive load. Extending this perspective, our findings illustrate how virtual comedians similarly engaged with the absence of nonverbal cues, not as a deficit but as a challenge to be worked with. By turning limitations into opportunities, they developed strategies to sustain performance and crafted what we call \textit{punchlines unbound}, expressions of creativity directed toward distributed audiences.

Meanwhile, our analysis revealed that the expressive repertoire of virtual comedians was deeply influenced by their technical proficiency (Sub-section~\ref{subsubsec:4-1-3}). 
Performers with such technical expertise were able to enhance their acts with elaborate gimmicks, while those without these skills, particularly professional comedians transitioning from physical stages, frequently expressed frustration at their inability to realize intended effects. 
This uneven distribution of technical knowledge also influenced how virtual comedians turn technical issues into performance opportunities. Disconnections, avatar glitches, and other breakdowns were often reinterpreted on the fly as material for improvisation, highlighting how technical issues themselves could become a resource for comedy. In other words, succeeding as a compelling comedian in social VR required not only comic sensibility and creativity but also a degree of technical skill.

\paragraph{\added{Design Implication 1}}
The technological implication drawn from these findings is not so much the introduction of entirely new platform-level features or toolkits, but rather the need to support performers who lack technical skills or are unfamiliar with social VR.
For example, applications like VRoid~\cite{isozaki2021vroid}, which allow users to create anime-like 3D characters and performance gimmicks through a simple GUI, are already widely used in social VR platforms. We see their potential to assist novice performers in live settings in social VR.
\subsection{Strategies and Challenges in Avatar-mediated Audience-Performer Interaction (RQ2)}
Performer–audience interaction in social VR comedy was constrained by the limited availability of spontaneous nonverbal cues (Section~\ref{subsubsec:4-2-1}). Our interviews revealed that current HMDs and controllers track only coarse head and hand movements, making it difficult for audiences to express subtle facial expressions or micro-gestures. As a result, performers struggled to interpret whether laughter or amusement was genuinely occurring. This limitation was amplified by the widespread use of non-humanoid avatars among audiences, which often lacked facial features or body parts that could convey smiling, clapping, or other gestures of appreciation. Prior research has further shown that text input in VR is cumbersome~\cite{doug2002textinputvr,mcgill2015textinput}, and our interviews confirmed that audiences found it difficult to provide immediate feedback through typing during live performances. Together, these factors created a communicative gap where audiences wanted to engage, but the available channels were too limited or too slow.

To bridge this gap, audiences developed distinctive interaction practices in social VR comedy (Section~\ref{subsubsec:4-2-1}). They used emoji not only for generic reactions (e.g., smile or clapping icons) but also as context-specific responses tied directly to the content of a joke. For instance, audiences sent a wallet emoji in response to jokes about money.
\added{Audiences used emojis to explicitly demonstrate their understanding of the joke and their amusement.}

From the performers' perspective, avatar-mediated interaction introduced a different set of challenges (Section~\ref{subsubsec:4-2-2}). A first issue was temporal delay. Comedy relies heavily on timing, and performers often pause briefly to ``wait for laughter'', a well-established technique in both theater and stand-up comedy. 
In social VR, network latency and audio jitter meant that laughter often arrived seconds after the punchline, making it difficult to judge whether a joke had landed. This delay undermined the effectiveness of timing strategies that are important to live comedy. 
A second issue was the audience's intentional feedback. While emoji reactions enabled diverse and contextually rich responses, they required deliberate selection and action from the audience. This made reactions slower and more effortful than spontaneous laughter or gestures~\cite{oh2016avatarsmile}. 

\paragraph{\added{Design Implication 2}} These challenges suggest two complementary directions for design interventions.  
First, capturing and transmitting natural reactions could help restore the immediacy of audience feedback.  
For example, systems that unobtrusively detect and share micro-expressions~\cite{zhang2023facialexpressionvr} or subtle physiological signals~\cite{robinson2022chathasnochill,han2023pshisiology} may enable performers to regain a sense of timing similar to that of live comedy, provided that privacy concerns are carefully addressed. 
By transmitting these subtle audience reactions in real-time, such systems could give performers earlier cues about whether a joke has landed, reducing the uncertainty caused by network delays.
Second, streamlining intentional feedback is equally important.  
Interfaces that reduce the effort required to select and send emoji~\cite{chiang2024sharingemoji} or avatar expressions~\cite{ohara2025cscw} could minimize delays and allow audiences to react more intuitively.  
Prior work on lightweight reaction mechanisms in collaborative settings~\cite{maeda2022calmresponse,yuill2012collaboration} has shown that such designs can support shared understanding, and applying similar principles to social VR performances may enable more fluid and accessible audience feedback.
In addition, recent advances in generative AI could be leveraged to recommend context-dependent reactions, thereby reducing the cognitive load associated with intentional selection.
Together, these approaches point toward systems that can make audience responses both easier to produce and more meaningful for live performers in social VR.

\subsection{\added{Community Practice for Virtual Live Performance (RQ3)}}
To sustain live performances, as our example of comedy in social VR, community-driven moderation becomes essential because institutional infrastructures in physical venues are absent (Section~\ref{subsubsec:4-3-1}).  
Virtual comedians frequently pointed out that behaviors such as excessive audience speech and trolling disrupted the flow of their performances.
Social VR platforms allow physical freedom, minimal spatial constraints, and strong anonymity \cite{blackwell2019harassment,mcveigh2019vrsocialinteraction}.
These features create an environment where audience members can participate more actively than in physical theaters.  
For example, in our study, audience members sometimes spoke at length or interjected loudly during punchlines, which comedians noted was especially disruptive in comedy because it broke the timing and rhythm of jokes.
While these can potentially be sources for improvisation, they also increase the risk of disrupting the performance.


Prior research on moderation in social VR has largely emphasized the prevention of harassment and personal attacks~\cite{freenan2022harassment,schulenverg2023harassment,blackwell2019harassment}.  
For example, platforms commonly provide blocking~\cite{chen2025socialvrblocking} and vote-kicking~\cite{chen2025moderation}, and reputation-based systems~\cite{chen2024trustsystem}.  
Although these were also observed in our study, we found that moderation in VR comedy involves more than just preventing misconduct.
Rather than only protecting individuals from harmful behavior, it was directed toward preserving the performance context itself.  
Some disruptions were not motivated by malice but stemmed from audience members' lack of awareness of community norms, such as entering the stage area or speaking over performers unintentionally.
\added{Therefore, rule-based moderation in physical performance was not always effective in the technology-mediated performance setting.}  
Instead, participants adopted informal practices, including sharing knowledge about etiquette, assigning ad-hoc responsibilities to experienced members, and collectively intervening when performances were at risk.

These practices can be understood through the concept of a community of practice~\cite{hoadley2005communities}.  
Participants accumulated shared norms and practical knowledge through ongoing engagement. 
Moderation became a collaborative process that evolved through the community's repeated experiences of managing disruptions.


\paragraph{\added{Design Implication 3}}
Our findings suggest that sustaining comedy performances in social VR cannot be achieved solely through platform-level controls.
Location-based or role-based permission systems, such as those in Cluster\footnote{Cluster allows event organizers to assign roles (e.g., Staff, Guest) and control participant permissions such as microphone use and spatial movement.\url{https://help.cluster.mu/hc/en-us/articles/360036092291-Creating-an-Event}}, can prevent interruptions, but they also risk undermining the openness and improvisational qualities that define social VR comedy.
If pushed too far, such designs may drift toward models of online video live streaming, where performers broadcast to passive audiences rather than engaging in the shared liveness and mutual presence that make social VR distinctive.
Similarly, trust-based moderation systems, such as VRChat's Trust Rank, also have limitations.
As Liu et al.~\cite{chen2024trustsystem} point out, reputation mechanisms can create boundaries that exclude newcomers from fully participating in the community.
This is especially problematic for live comedy, which often relies on spontaneous audience participation.


Instead, our findings show that moderation in social VR emerges through the cultivation of community norms and the sharing of knowledge across platforms.
Alcala et al.~\cite{alcala2023socialvrdiscord} describe this as a ``stage and theater'' relation between VR platforms and Discord.
We observed similar patterns in VRChat comedy, where performers and audiences used Discord to share information about disruptive users, pass down etiquette, and coordinate events.

\subsection{Limitation and Future Directions}
One limitation of this study is that the investigation was conducted within a Japanese-speaking VRChat community.
Although members performed across different platforms and events, the group remains relatively close-knit.
The community is still developing, and as social VR platforms continue to expand, newcomers with different levels of technical literacy and varying levels of shared cultural references may change how performances are produced and received.
Future research should examine how community growth and diversification affect practices such as improvisation, moderation, and the use of VR as a shared epistemic ground.

Another important limitation concerns cultural differences.
In our study, we observed relatively universal formats such as stand-up and sketch comedy, which suggests that some of the strategies we identified may be transferable across contexts.
At the same time, comedy styles and audience responses are strongly shaped by local norms.
For instance, British pantomime\footnote{https://en.wikipedia.org/wiki/Pantomime} may involve performance practices that differ from those we observed.
Focusing on culturally specific forms of comedy can therefore provide new insights into how avatar-mediated performance adapts to different traditions. Future comparative studies across cultural settings would help identify which practices are widely shared and which are uniquely shaped by local contexts.

Finally, this study relied on interviews with performers, which means that audience experiences were explored only indirectly.
A deeper investigation of how audiences perceive and participate in social VR comedy would be valuable, and this perspective would also enrich the study of live performance more broadly. 
In addition, our analysis depended mainly on interview data. Future work could combine interviews with other methods such as participatory design or prototyping, which may help translate the challenges identified in this study, such as constraints on audience feedback and the difficulties of moderation, into more concrete directions for technical intervention.


\section{CONCLUSION}
This study examined how virtual comedians and audiences in social VR adapt to the constraints and features of avatar-mediated performance.
Our findings showed that performers reinterpreted limited nonverbal expressiveness into creative staging strategies, audiences developed distinctive feedback practices such as context-adaptive emoji use, and communities enacted forms of self-governance to manage disruptions while drawing on VR itself as shared cultural ground.
These dynamics reveal that social VR comedy transforms traditional modes of performance: constraints become resources for creativity, yet challenges remain in sustaining immediate feedback and scaling shared knowledge.
While these tensions highlight the fragility of performance in expanding communities, they also point to opportunities for design interventions.
Future work should examine how these practices evolve with more diverse audiences and larger scales, and future systems should support performers and communities through accessible expressive tools, mechanisms for capturing or streamlining audience reactions, and lightweight moderation features that preserve both creative freedom and cultural continuity.  

\begin{acks}
This work was supported by JST COI-NEXT (JPMJPF2201), JST Moonshot (JPMJMS2013) and JST SPRING (JPMJSP2108).
\end{acks}

\bibliographystyle{ACM-Reference-Format}
\bibliography{main_ref}

\appendix

\section{APPENDIX}
\subsection{Comedy Styles}
\label{appendix:comedystyle}
\begin{figure*}[htbp]
  \centering
  \begin{subfigure}[t]{0.24\linewidth}
    \centering
    \includegraphics[height=4cm]{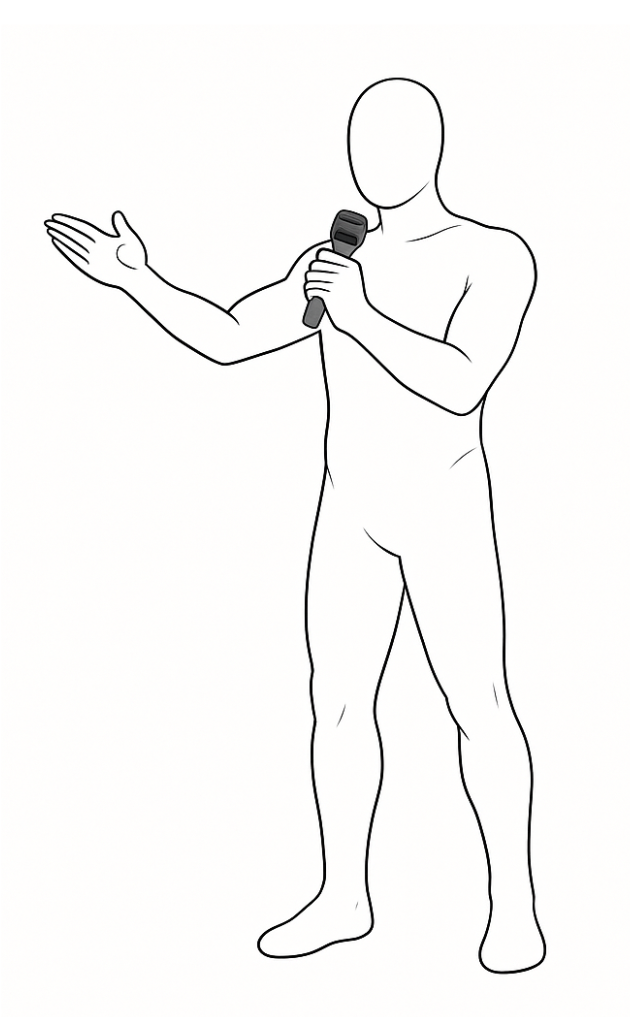}
    \caption{Stand-up}
    \label{fig:standup}
  \end{subfigure}
  \hfill
  \begin{subfigure}[t]{0.24\linewidth}
    \centering
    \includegraphics[height=4cm]{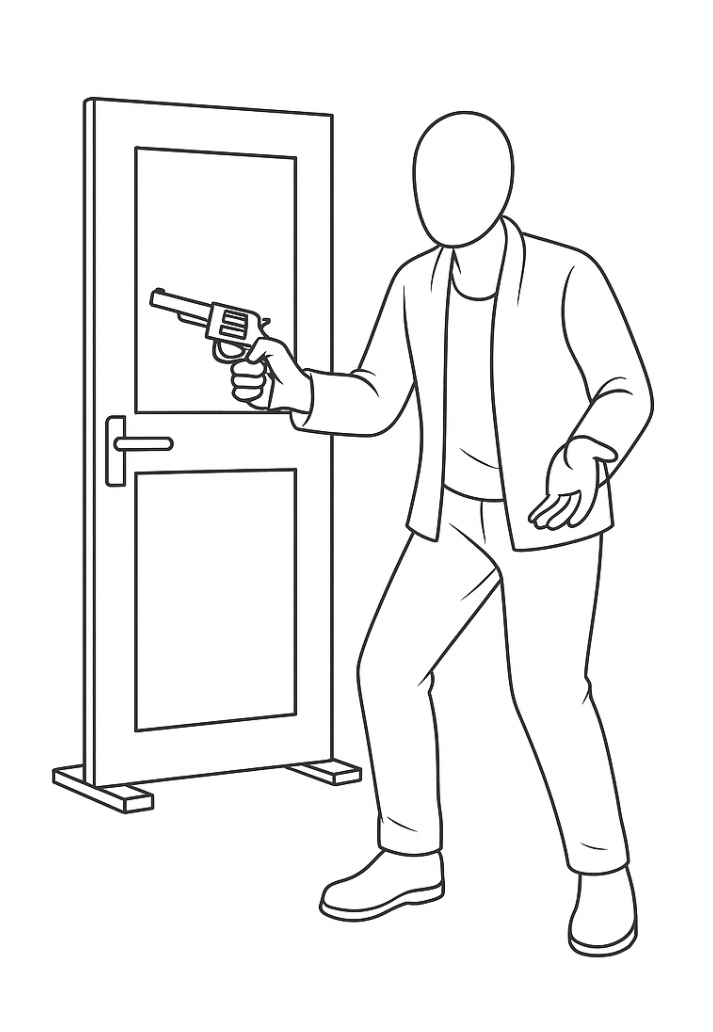}
    \caption{Solo-sketch}
    \label{fig:solosketch}
  \end{subfigure}
  \hfill
  \begin{subfigure}[t]{0.24\linewidth}
    \centering
    \includegraphics[height=4cm]{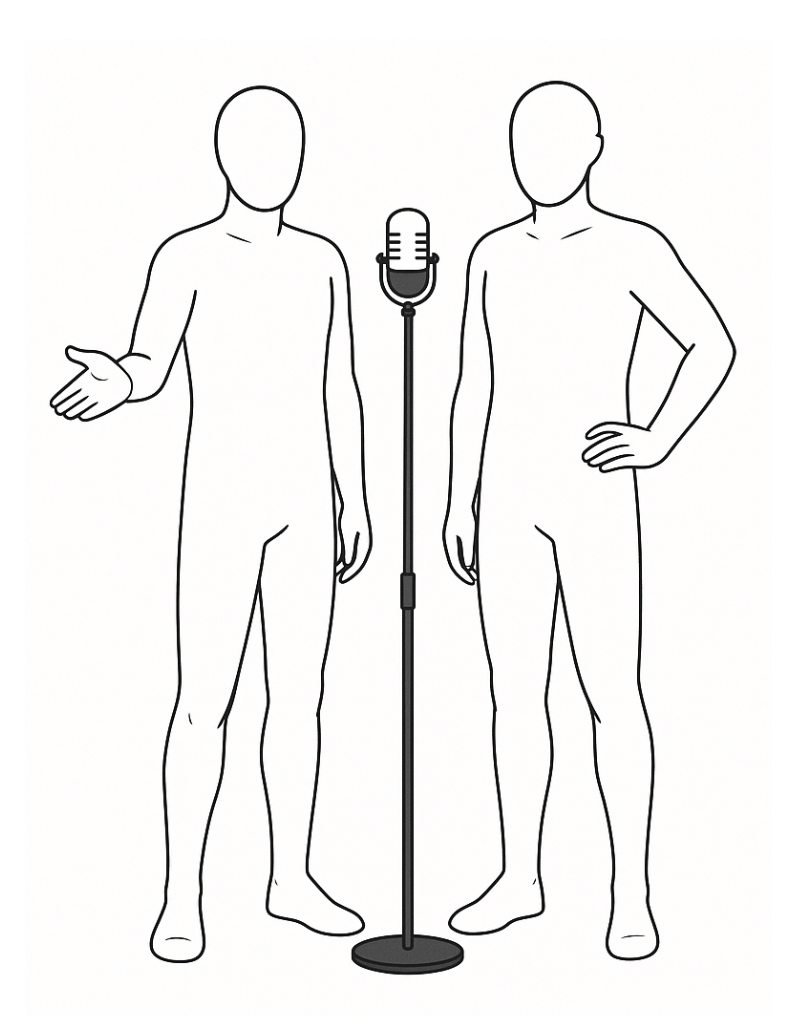}
    \caption{Manzai}
    \label{fig:manzai}
  \end{subfigure}
  \hfill
  \begin{subfigure}[t]{0.24\linewidth}
    \centering
    \includegraphics[height=4cm]{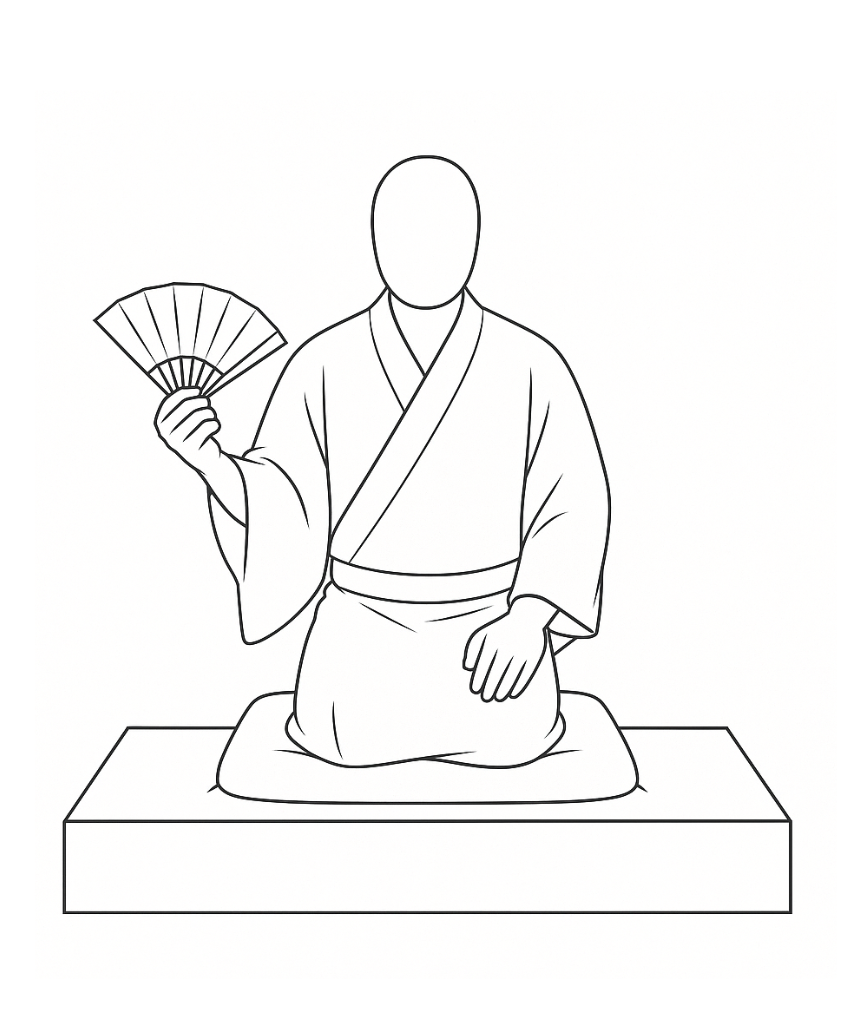}
    \caption{Rakugo}
    \label{fig:rakugo}
  \end{subfigure}
  \caption{Illustrations of the four comedy styles observed in this study.}
  \label{fig:comedystyle}
\end{figure*}

Figure~\ref{fig:comedystyle} illustrates the four comedy styles observed in this study: Stand-up, Solo-sketch, Rakugo, and Manzai.
\textbf{Stand-up} involves a single comedian delivering a spoken monologue using a microphone, typically focusing on verbal storytelling or observational humor.
\textbf{Solo-sketch} features one performer enacting a short narrative scenario, sometimes incorporating props or visual effects to enhance the comedic setup.
\textbf{Manzai} is a two-person comedic dialogue characterized by rapid back-and-forth interaction, generally structured around the roles of ``straight person'' (tsukkomi) and ``funny person'' (boke)~\cite{katayama2006manzai,katayama2008manzai}.
\textbf{Rakugo} is a traditional form of seated storytelling in which a single performer portrays multiple characters within a scripted narrative~\cite{sakai1981rakugo}.

\subsection{Interview Guide}
\label{appendix:interviewguide}

\begin{itemize}
  \item \textbf{RQ1: Performers' Strategies in Social VR}
    \begin{itemize}
      \item You use a specific avatar (e.g., Panda) for your performances. How did you decide on this avatar when you first started performing?
      \item (After checking whether the interview avatar is the same) Do you prefer to use the same avatar as in your everyday social VR use, or a different one? Why?
      \item Are there situations where the current avatar you are using has helped you in your comedy performances?
      \item When you perform in VRChat, do you see yourself as the same person as when you use VRChat in daily life?
      \item During your performance, do you use an HMD or a PC? Has this been an advantage or a constraint in delivering jokes?
      \item How do you use tracking during performances? Has this been an advantage or a constraint for your routines?
      \item Have you ever created jokes that make use of the characteristics of virtual space or virtual props?
      \item Do you draw on the comedy you have seen in the physical world when creating your own material?
      \item (If comfortable) Do you have experience performing in the physical world? If so, what differences have you noticed in creating jokes between physical and virtual contexts?
      \item Among your performances so far, which ones were the most memorable for you? Why?
      \item When you perform, do you prefer to invite real-world friends or VRChat friends?
    \end{itemize}

  \item \textbf{RQ2: Performer–Audience Interaction in Social VR}
    \begin{itemize}
      \item During your performances, can you tell when the audience is enjoying your act?
      \item In what forms do audiences express their laughter?
      \item Does performing in VR create constraints in your interaction with the audience?
      \item (If applicable) For those who have real-world performance experience: what are the main differences between live and virtual audience interactions?
    \end{itemize}

  \item \textbf{RQ3: Cultural Practices and Community Norms}
    \begin{itemize}
      \item Can you describe the flow of preparing and hosting a comedy show in VR?
      \item Are there any rules or cautions you give to audiences before the show? Were these based on actual incidents that happened in past performances?
      \item What was the most memorable event that occurred during a performance? Why?
      \item Can you recall any particularly memorable audience behavior?
      \item Do you think the style of comedy viewing in VR is the same as in physical comedy theaters?
      \item What are the unique advantages of performing comedy in VR compared to physical settings?
    \end{itemize}
\end{itemize}

\subsection{Emoji Reactions in VRChat}
\label{appendix:emoji}
\begin{figure}[htbp]
    \centering
    \includegraphics[height=5cm]{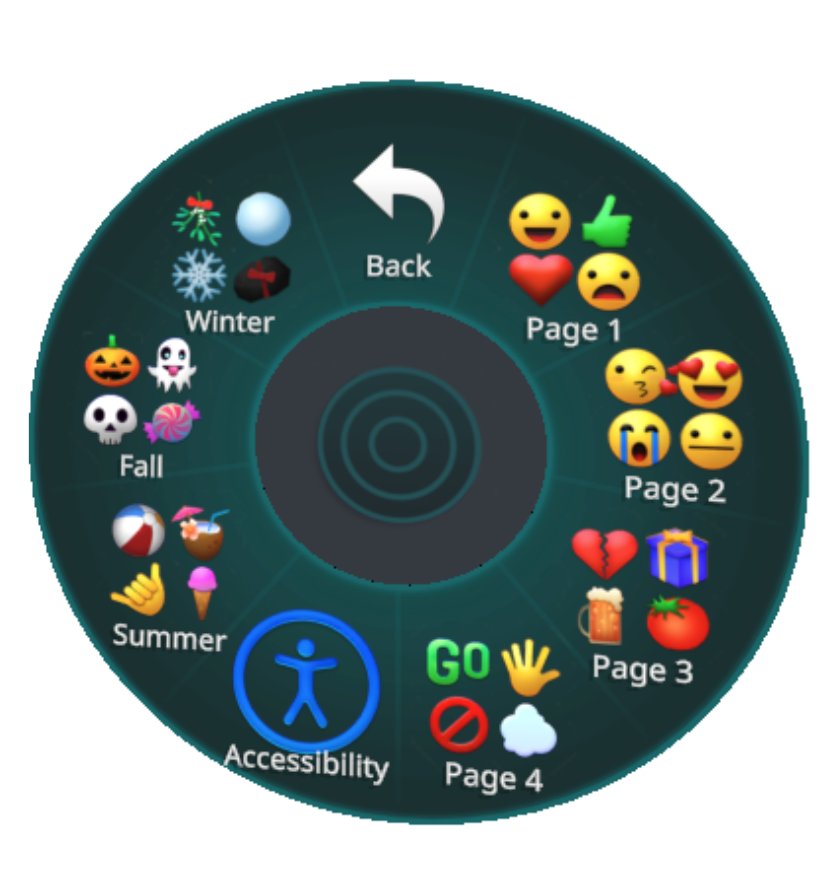}
    \caption{Radial menu interface in VRChat used to trigger emoji reactions. Source: VRChat Documentation Wiki~\url{https://wiki.vrchat.com/wiki/Action_Menu}, licensed under CC BY-SA.}
    \label{fig:emojis}
\end{figure}

Figure~\ref{fig:emojis} shows the radial menu interface in VRChat, which users employ to trigger emoji reactions.
By selecting an emoji from this menu, audience members can display floating symbols above their avatars in the shared space.
These reactions are ephemeral and highly visible, and they often substitute for subtle nonverbal feedback such as quiet laughter or applause.

\end{document}